\documentclass[a4paper,10pt,twoside]{cpc-hepnp}

\usepackage{multicol}
\usepackage{graphicx}
\usepackage{booktabs}
\usepackage{amssymb,bm,mathrsfs,bbm,amscd}
\usepackage[tbtags]{amsmath}
\usepackage{lastpage}
\usepackage{CJK}

\begin{document}

\title{Particle-number conserving analysis for the 2-quasiparticle and
high-$K$ multi-quasiparticle states in doubly-odd ${}^{174, 176}$Lu
\thanks{Supported by NSFC (Grant No. 10875157 and 10979066),
MOST (973 Project 2007CB815000), and CAS (Grant No. KJCX2-EW-N01 and KJCX2-YW-N32).}
}

\begin{CJK*}{GBK}{song}
\author{
LI Bing-Huan$^{1}$ \quad
ZHANG Zhen-Hua$^{2;1)}$\email{zhzhang@itp.ac.cn} \quad
LEI Yi-An$^{1}$
}

\maketitle
\end{CJK*}

\address{
1~(State Key Lab of Nuclear Physics and Technology, School of Physics,
   Peking University, Beijing 100871, China)\\
2~(State Key Laboratory of Theoretical Physics,
   Institute of Theoretical Physics, \\
   Chinese Academy of Sciences, Beijing 100190, China)\\
}

\begin{abstract}
Two-quasiparticle bands and low-lying excited high-$K$ four-, six-, and
eight-quasiparticle bands in the doubly-odd ${}^{174, 176}$Lu are analyzed
by using the cranked shell model (CSM) with the pairing correlations
treated by a particle-number conserving (PNC) method,
in which the blocking effects are taken into account exactly.
The proton and neutron Nilsson level schemes for ${}^{174, 176}$Lu are taken
from the adjacent odd-$A$ Lu and Hf isotopes, which are adopted to reproduce the
experimental bandhead energies of the one-quasiproton and one-quasineutron
bands of these odd-$A$ Lu and Hf nuclei, respectively.
Once the quasiparticle configurations are determined, the experimental bandhead
energies and the moments of inertia of these two- and multi-quasiparticle
bands are well reproduced by PNC-CSM calculations.
The Coriolis mixing of the low-$K$ ($K=|\Omega_1-\Omega_2|$) two-quasiparticle
band of the Gallagher-Moszkowski doublet with one nucleon in the $\Omega = 1/2$
orbital is analyzed.
\end{abstract}



\footnotetext[0]{\hspace*{-2em}\small\centerline{\thepage\ --- \pageref{LastPage}}}%

\begin{multicols}{2}

\section{Introduction}
The level structure of doubly-odd deformed nuclei is among the most complex
topics in nuclear physics because of the complexity of level structure
associated with contributions from both valence protons and neutrons.
However, they often provide a wealth of nuclear structure phenomena.
The rotational bands of deformed doubly-odd nuclei exhibit several unusual
features and anomalies such as Gallagher-Moszkowski (GM)
splittings~\cite{Gallagher1958_PR0111-1282},
Newby shifts~\cite{Newby1962_PR0125-2063},
chiral structure~\cite{Frauendorf1997_NPA617-131},
signature inversion~\cite{Kreiner1979_PRL43-1150, Kreiner1980_JPG6-L13}, etc.,
which have been investigated both from experimental and theoretical
sides~\cite{Boisson1976_PR26-99, Bengtsson1984_NPA415-189,
Frisk1988_ZPA330-241, Hara1991_NPA531-221,
Liu1995_PRC52-2514, Liu1996_PRC54-719, Bark1997_PLB406-193, Jain1998_RMP70-843,
Xu2000_NPA669-119, Starosta2001_PRL86-971, Frauendorf2001_RMP73-463,
Olbratowski2004_PRL93-052501, Meng2010_JPG37-064025}.
A striking feature of axially symmetric deformed rare-earth nuclei with
$A\sim180$ is the observation of a large number of long-lived
isomers~\cite{Walker1999_Nature399-35}.
Especially many very long-lived isomers are observed in the doubly-odd nuclei.
The most famous isomer $K^\pi = 9^-$ $(9/2^-[514] \otimes 9/2^+[624])$ in $^{180}$Ta
has a half-life longer than $1.2 \times 10^{15}$ years~\cite{Cumming1985_PRC31-1494},
while the ground state $K^\pi = 1^+$ $(7/2^+[404] \otimes 9/2^+[624])$
has a half-life about a few hours.
These high-$K$ bands arise from the coupling of a few
orbitals near both proton and neutron Fermi surfaces, with large
angular momentum projections $\Omega_i$ on the nuclear symmetry axes
($K=\sum_{i}\Omega_i$); e.g., the proton orbitals $\pi7/2^+[404] (g_{7/2})$,
$\pi9/2^-[514] (h_{11/2})$, $\pi5/2^+[402] (d_{5/2})$,
and the neutron orbitals $\nu7/2^+[633] (i_{13/2})$,
$\nu7/2^-[514] (h_{9/2})$, $\nu9/2^+[624] (i_{13/2})$, etc.

Most of the earlier studies in this region have focused on the neutron deficient nuclei.
With the development of experimental techniques,
more and more measurements have been done for the neutron rich nuclei.
As for the neutron rich doubly-odd Lu nuclei, many experiments have been performed
not only for the 2-quasiparticle (qp) structures, but also for the high-$K$ multi-qp
isomers~\cite{Oneil1972_NPA195-207, Bruder1987_NPA467-1, Bruder1987_NPA474-518,
Drissi1990_NPA512-413, Klay1991_PRC44-2801, McGoram2000_PRC62-031303R,
Dracoulis2006_PRL97-122501, Kondev2009_PRC80-014304, Dracoulis2010_PRC81-011301R}.

It is well known that pairing correlations are very important in the low angular
momentum region, where they are manifested by reducing the  moments of inertia (MOI's)
of the rigid-body estimation~\cite{Bohr1958_PR0110-936}. Due to the blocking effects,
the MOI's of 1-qp bands in odd-$A$ nuclei are usually larger than those of the
ground state bands (gsb's) in adjacent even-even nuclei~\cite{Bohr1975_Book}.
The blocking effects on MOI's of multi-qp bands are even
more important~\cite{Zeng2002_PRC65-044307}.
In Ref.~\cite{Zhang2009_PRC80-034313}, the systematically observed low-lying
high-$K$ multi-qp bands in even-even and odd-$A$ Hf and Lu
isotopes ($170\leq A \leq 178$)~\cite{Singh2006_ADNDT92-1} are analyzed using
the cranked shell model (CSM) with pairing correlations treated by a particle-number
conserving (PNC) method~\cite{Zeng1983_NPA405-1, Zeng1994_PRC50-1388}.
Once an appropriate Nilsson level scheme near the Fermi surface for a given
nucleus is adopted to reproduce the experimental bandhead energies of 1-qp bands
in an odd-$A$ nucleus, the experimental MOI's of these 1-qp bands
and various kinds of low-lying high-$K$ multi-qp bands can  be reproduced well by
the PNC-CSM calculations.
In contrary to the conventional Bardeen-Cooper-Schrieffer (BCS) or
Hartree-Fock-Bogolyubov (HFB) approaches, in the PNC method, the  Hamiltonian is
solved directly in a truncated Fock-space~\cite{Wu1989_PRC39-666}.
So the particle-number is conserved and the Pauli blocking effects are taken into
account exactly. Note that the PNC scheme has been used both in relativistic
and nonrelativistic mean field models~\cite{Meng2006_FPC1-38, Pillet2002_NPA697-141},
in which the single-particle states are calculated from the self-consistent
mean field potentials instead of the Nilsson potential. The PNC-CSM also have been
used to investigate the heaviest nuclei with proton number $Z \approx 100$
~\cite{He2009_NPA817-45, Zhang2011_PRC83-011304R, Zhang2012_PRC85-014324}.

In this paper, using the Nilsson level schemes of the odd-$A$ Lu and Hf isotopes
adopted in Ref.~\cite{Zhang2009_PRC80-034313}, we present the results of the doubly-odd
${}^{174, 176}$Lu from PNC-CSM calculations. The paper is organized as follows.
A brief introduction of the PNC treatment of pairing correlations within
the CSM is presented in Sec.~2.
The PNC-CSM calculations of the excitation energies and MOI's for states
in ${}^{174, 176}$Lu are presented in Sec.~3.
A brief summary is given in Sec.~4.

\section{A brief introduction to PNC-CSM}{\label{Sec:PNC}}

The CSM Hamiltonian of an axially symmetric nucleus in the
rotating frame is~\cite{Zeng1994_PRC50-1388, Xin2000_PRC62-067303},
\begin{equation}
 H_\mathrm{CSM} = H_0                    + H_\mathrm{P}
                = H_{\rm Nil}-\omega J_x + H_\mathrm{P} \ ,
 \label{eq:H_CSM}
\end{equation}
where $H_{\rm Nil}$ is the Nilsson Hamiltonian~\cite{Nilsson1969_NPA131-1},
$-\omega J_x$ is the Coriolis interaction with the cranking frequency
$\omega$ about the $x$ axis (perpendicular to the nuclear symmetry $z$ axis).
$H_{\rm P}$ is the monopole pairing interaction
\begin{equation}
 H_{\rm P} =
  -G \sum_{\xi\eta} a^\dag_{\xi} a^\dag_{\bar{\xi}}
                        a_{\bar{\eta}} a_{\eta} \ ,
\end{equation}
where $\bar{\xi}$ ($\bar{\eta}$) labels the time-reversed state of a
Nilsson state $\xi$ ($\eta$), and $G$ is the effective strengths of
monopole pairing interaction, which is determined by the experimental
odd-even differences in nuclear binding energies.

Instead of the usual single-particle level truncation used in common
shell model calculations, a cranked many-particle configuration
(CMPC) truncation (Fock space truncation) is adopted which is crucial
to make the PNC calculations for low-lying excited states both workable
and sufficiently accurate~\cite{Wu1989_PRC39-666, Molique1997_PRC56-1795}.
An eigenstate of $H_\mathrm{CSM}$ can be written as
\begin{equation}
 | \psi \rangle = \sum_{i} C_i | i \rangle , \qquad (C_i \; \textrm{real}) \ ,
 \label{eq:eigenstate}
\end{equation}
where $| i \rangle$ is an eigenstate of the one-body operator $H_0$, i.e., a CMPC.
By diagonalizing $H_\mathrm{CSM}$ in a sufficiently large CMPC space,
sufficiently accurate solutions for low-lying excited eigenstates of
$H_\mathrm{CSM}$ can be obtained.

The angular momentum alignment of state $|\psi\rangle$ is,
\begin{equation}
 \langle \psi | J_x | \psi \rangle =
 \sum_i C_i^2 \langle i | J_x | i \rangle +
 2\sum_{i<j}C_i C_j \langle i | J_x | j \rangle \ ,
 \label{eq:jx}
\end{equation}
and the kinematic MOI of state $| \psi \rangle$ is
\begin{equation}
 J^{(1)} = \frac{1}{\omega} \langle\psi | J_x | \psi \rangle \ .
\end{equation}
The occupation probability $n_\mu$ of the cranked orbital
$|\mu\rangle$ is $n_{\mu}=\sum_{i}|C_{i}|^{2}P_{i\mu}$, where
$P_{i\mu}=1$ if $|\mu\rangle$ is occupied in $|i\rangle$, and
$P_{i\mu}=0$ otherwise.

In the following PNC-CSM calculations, the Hamiltonian~(\ref{eq:H_CSM}) is
diagonalized in a sufficiently large CMPC space to obtain the low-lying
excited eigenstates $|\psi\rangle$'s.
The dimension of the CMPC space is about 700 for protons and 800 for neutrons.
As we are only interested in the yrast and low-lying
excited states, the number of the important CMPC's involved (weight
$>1\%$) is not very large (usually $<20$) and almost all the CMPC's
with weight $>0.1\%$ are counted in, so the solutions to the
low-lying excited states are quite accurate.

We note that because $R_{x}(\pi)=e^{-i \pi J_x}$, $[J_{x}, J_{z}]\neq0$,
the signature scheme invalidates the quantum number $K$.
It has been pointed out that~\cite{Zeng1994_PRC50-1388, Wu1990_PRC41-1822},
although $[J_{x}, J_{z}] \neq 0$, we have $[R_{x}(\pi), J_{z}^2] = 0$.
Thus we can construct simultaneous eigenstates of ($R_{x}(\pi), J_{z}^2$).
Each CMPC $|i\rangle$ in Eq.~(\ref{eq:eigenstate}) is chosen as a
simultaneous eigenstate of ($H_0, J_{z}^2$).
However, $K$ is still commonly used as a convenient quantum number to
label rotational bands of deformed nuclei.
For the gsb of an even-even nucleus, parity $\pi=+$,
$\alpha=0$, $K=0$, $I=0, 2, 4,\cdots$.
For an odd-$A$ nucleus, if the Nilsson orbital 1 is blocked by an unpaired nucleon,
we have two sequences of rotational levels with $\pi=\pi_1$, $K=\Omega_1$,
$\alpha=\pm1/2$, $I\geq K$.
For a 2-qp band in an doubly-odd nucleus, when the proton and neutron
Nilsson orbitals 1 and 2 are blocked by two unpaired nucleons,
we have four sequences of rotational levels, $\pi=\pi_1\pi_2$,
$K=|\Omega_1\pm\Omega_2|$, $\alpha=0, 1$, $I\geq K$.
The situation is similar for pair-broken (multi-qp) bands, etc.

\section{Results and discussions}{\label{results}}
\subsection{Parameters}
Comparison of the PNC-CSM calculations on excitation energies and MOI's for the low-lying
states in ${}^{174, 176}$Lu with the experimental values is shown in this section.
The deformation parameters $\varepsilon_2 = 0.262, \varepsilon_4 = 0.0402$
($\varepsilon_2 = 0.260, \varepsilon_4 = 0.0512$) for ${}^{174}$Lu (${}^{176}$Lu)
are taken from Lund systematics~\cite{Bengtsson1986_ADNDT35-15},
i.e., an average of the neighboring even-even Yb and Hf isotopes.
The proton and neutron Nilsson level schemes of ${}^{174}$Lu (${}^{176}$Lu) are taken
from those of ${}^{173}$Lu and ${}^{175}$Hf (${}^{177}$Lu and ${}^{177}$Hf)
in Ref.~\cite{Zhang2009_PRC80-034313}, which are adopted to reproduced the bandhead
energies of low-lying 1-quasiproton and 1-quasineutron bands in
${}^{173}$Lu and ${}^{175}$Hf (${}^{177}$Lu and ${}^{177}$Hf).

The effective pairing strengths can be determined by the odd-even differences in
nuclear binding energies, and are connected with the dimension of the truncated CMPC space.
The CMPC space for ${}^{174, 176}$Lu is constructed in the proton $N = 4, 5$ shells
and the neutron $N = 5, 6$ shells.
The dimensions of the CMPC space are about 700 for protons and 800 for neutrons
in our calculation. The truncated CMPC's energies are about $0.8\hbar\omega_0$
for protons and $0.7\hbar\omega_0$ for neutrons, where $\hbar\omega_0 = 41A^{-1/3}$~MeV.
The effective pairing interaction strengths for
${}^{174}$Lu (${}^{176}$Lu) are $G_p = 0.33$~MeV (0.30~MeV) for protons and
$G_n = 0.30$~MeV (0.28~MeV) for neutrons, respectively.
The stability of the PNC-CSM calculation results against the change
of the dimension of the CMPC space has been investigated in Refs.~\cite{Zeng1994_PRC50-1388,
Liu2002_PRC66-024320, Molique1997_PRC56-1795, Zhang2012_PRC85-014324}.
In the present calculations, almost all the CMPC's with weight $>0.1\%$ are taken
into account, so the solutions to the low-lying excited states are accurate enough.
A larger CMPC space with renormalized pairing strengths gives essentially the same results.

When an unpaired proton and an unpaired neutron in a
deformed doubly-odd nucleus are coupled, the projections of their total
angular momentum on the nuclear symmetry axis, $\Omega_p$ and $\Omega_n$, can
produce two states with $K_> =\Omega_p+ \Omega_n$ and $K_< =|\Omega_p - \Omega_n|$.
They follow the GM coupling rules~\cite{Gallagher1958_PR0111-1282}:
\begin{eqnarray}
 K_> &=& |\Omega_p + \Omega_n|, \ \text{if} \ \Omega_p=\Lambda_p \pm \frac{1}{2} \
                                 \text{and} \ \Omega_n=\Lambda_n \pm \frac{1}{2} \ ,
\nonumber\\
 K_< &=& |\Omega_p - \Omega_n|, \ \text{if} \ \Omega_p=\Lambda_p \pm \frac{1}{2} \
                                 \text{and} \ \Omega_n=\Lambda_n \mp \frac{1}{2} \ .
\nonumber
\end{eqnarray}
Note that the neutron-proton residual interaction is not included in our calculation,
so we can not get the GM doublets splittings.

\subsection{${}^{174}$Lu}

\begin{center}
\includegraphics[width=8cm]{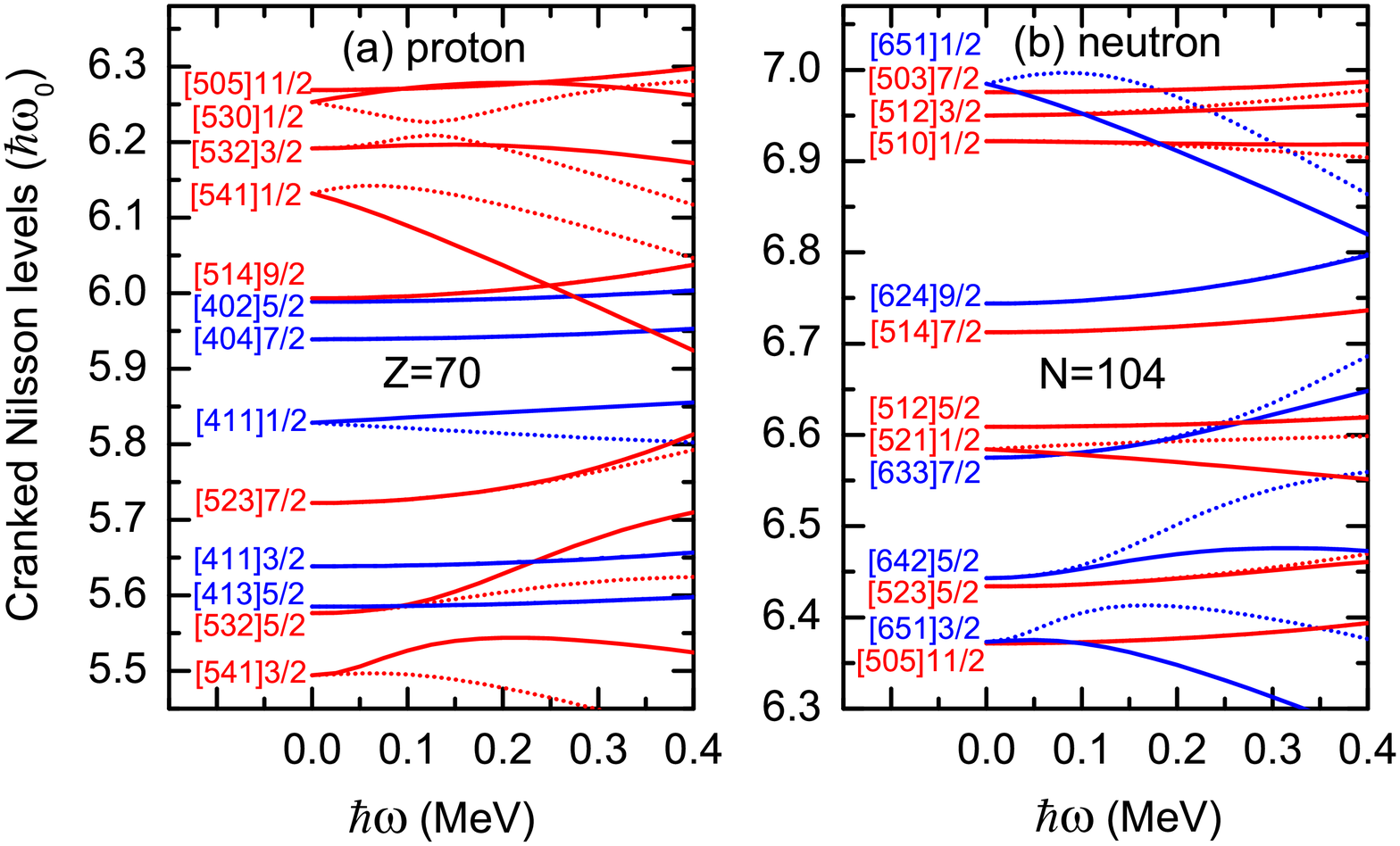}
\figcaption{ \label{fig:174LuNilsson}
(Color online) The cranked Nilsson levels near
the Fermi surface of ${}^{174}$Lu for (a) protons and (b) neutrons.
The deformation parameters ($\varepsilon_2 = 0.262, \varepsilon_4 = 0.0402$)
are taken from Lund systematics~\citep{Bengtsson1986_ADNDT35-15},
i.e., an average of the neighboring even-even Yb and Hf isotopes.
The proton (neutron) Nilsson level scheme of ${}^{174}$Lu is taken from that of
${}^{173}$Lu (${}^{175}$Hf) in Ref.~\citep{Zhang2009_PRC80-034313},
which is adopted to reproduced the bandhead energies of
low-lying 1-quasiproton (1-quasineutron) bands of ${}^{173}$Lu (${}^{175}$Hf).
The positive (negative) parity levels are denoted by the blue (red) lines.
The signature $\alpha= +1/2$ ($\alpha = -1/2$) levels are denoted
by the solid (dotted) lines.
}
\end{center}

The cranked Nilsson levels near the Fermi surface of ${}^{174}$Lu are given
in Fig.~\ref{fig:174LuNilsson}.
The positive (negative) parity levels are denoted by the blue (red) lines.
The signature $\alpha= +1/2$ ($\alpha = -1/2$) levels are denoted
by the solid (dotted) lines. Figure~\ref{fig:174LuNilsson} shows that there exist a
proton sub-shell at $Z = 70$ and a neutron sub-shell at $N = 104$.

Using this Nilsson level scheme, first we calculate the bandhead energies of
2- and multi-qp states in $^{174}$Lu with pairing interaction included.
The results are shown in Table~\ref{tab:174Lu}.
Most of the data are well reproduced by the PNC-CSM calculations even without the
neutron-proton residual interaction, especially those low-lying high-$K$ multi-qp states.
The only obvious exception is $K^\pi = 2^+$ ($\pi1/2[541] \otimes \nu5/2^-[512]$).
In our calculation the energy of $\pi1/2[541]$ is too high.
It is very hard to reproduce the correct location of this orbital in the Nilsson level scheme.
For ${}^{174}$Lu, the last occupied proton Nilsson orbital is $\pi7/2^+[404]$ and the last
occupied neutron Nilsson orbital is $\nu5/2^-[512]$, which is consistent with the data.
The experimental values of these 2-qp bands indicate that the orbital $\nu7/2^+[633]$ is
more closer to the neutron Fermi surface than $\nu1/2^-[521]$, which is reversed in our
calculation.
We should note that, the level sequence in odd-$A$ Hf isotopes in the $170<A< 180$
mass region~\cite{Jain1990_RMP62-393} is consistent with what we used.
This level sequence inversion still needs to be investigated.
For the 4-, 6-, and 8-qp states, the possible configurations and the corresponding
bandhead energies are also shown. The results are quite similar to those multi-qp
blocking calculations in Ref.~\cite{Kondev2009_PRC80-014304}.
For the state at 4710~keV, the calculated energies of the two configuration assignments
$K^\pi=22^+$ and $K^\pi=22^-$ are both close to the data. So these two configurations
are all possible for the observed state with excitation energy 4710~keV,
which is consistent with the conclusion in Ref.~\cite{Kondev2009_PRC80-014304}.
Later on, the possible configuration assignments for $K^+ = 14^- ,15^+$, and $16^+$
will be analyzed according to the MOI's. For the calculations of the excitation
energies and the MOI's, the configuration assignments can be reasonably made.

\end{multicols}
\begin{center}
\tabcaption{ \label{tab:174Lu} The experimental and calculated 2- and multi-qp states in $^{174}$Lu.
The experimental values are taken from Refs~\citep{Bruder1987_NPA467-1, Bruder1987_NPA474-518,
Firestone1999, Dracoulis2006_PRL97-122501, Kondev2009_PRC80-014304}.
}
\footnotesize
\begin{tabular*}{135mm}{@{\extracolsep{\fill}}cccc}
\toprule
$K^\pi$      & Configuration$^{\rm a}$& $E_{\rm Exp}$ (keV)& $E_{\rm Cal}$ (keV)\\
\hline
$1^-, 6^-$   &  $\pi7/2^+ \otimes \nu5/2^- $               & 0, 171   & 0    \\
$2^+ $       &  $\pi1/2^- \otimes \nu5/2^- $               & 241      & 1330 \\
$0^+, 7^+$   &  $\pi7/2^+ \otimes \nu7/2^+ $               & 281, 431 & 258  \\
$4^-, 3^-$   &  $\pi7/2^+ \otimes \nu1/2^- $               & 365, 433 & 195  \\
$5^- $       &  $\pi5/2^+ \otimes \nu5/2^- $               & 456      & 370  \\
$7^+, 2^+ $  &  $\pi9/2^- \otimes \nu5/2^- $               & 531, 635 & 399  \\
$8^- $       &  $\pi9/2^- \otimes \nu7/2^+ $               & 772      & 657  \\
\hline
$13^+$       &  $\pi7/2^+ \otimes \nu^3 7/2^+,5/2^-,7/2^- $& 1856     & 1638 \\
$14^- $      &  $\pi9/2^- \otimes \nu^3 7/2^+,5/2^-,7/2^- $& 2063     & 2037 \\
$14^{-\rm b}$&  $\pi7/2^+ \otimes \nu^3 7/2^+,9/2^+,5/2^- $& 2063     & 1873 \\
$15^+    $   &  $\pi9/2^- \otimes \nu^3 7/2^+,9/2^+,5/2^- $& 2732     & 2274 \\
$15^{+\rm c}$&  $\pi^3 7/2^+,7/2^-,9/2^- \otimes \nu 7/2^+$& 2732     & 2490 \\
$16^+    $   &  $\pi9/2^- \otimes \nu^3 7/2^+,9/2^+,7/2^- $& 2876     & 2901 \\
$16^{+\rm d}$&  $\pi^3 7/2^+,7/2^-,9/2^- \otimes \nu 9/2^+$& 2876     & 2885 \\
$19^+    $   &  $\pi9/2^- \otimes
                 \nu^5 7/2^+,9/2^+,1/2^-,5/2^-,7/2^-$      & 3741     & 3629 \\
$21^+    $   &  $\pi^3 7/2^+,7/2^-,9/2^- \otimes
                 \nu^3 7/2^+,5/2^-,7/2^-$                  & 4069     & 4127 \\
$22^-    $   &  $\pi^3 7/2^+,7/2^-,9/2^- \otimes
                 \nu^3 7/2^+,9/2^+,5/2^-$                  & 4710     & 4362 \\
$22^{+\rm e}$&  $\pi^3 7/2^+,7/2^-,9/2^- \otimes
                 \nu^3 9/2^+,5/2^-,7/2^-$                  & 4710     & 4950 \\
$23^-    $   &  $\pi^3 7/2^+,7/2^-,9/2^- \otimes
                 \nu^3 7/2^+,9/2^+,7/2^-$                  & 5062     & 4991 \\
$26^-    $   &  $\pi^3 7/2^+,7/2^-,9/2^- \otimes
                 \nu^5 7/2^+,9/2^+ ,1/2^-,5/2^-,7/2^-$     & 5850     & 6119 \\
\bottomrule
\multicolumn{4}{l}{$^{\rm a}$ Protons ($\pi$):
$ 5/2^+: 5/2^+[402]; 7/2^+: 7/2^+[404]; 1/2^-: 1/2^-[541];  7/2^-: 7/2^-[523];
9/2^-: 9/2^-[514]$. }\\
\multicolumn{4}{l}{ Neutrons ($\nu$):
$1/2^-: 1/2^-[521]; 5/2^-: 5/2^-[512]; 7/2^-: 7/2^-[514];
7/2^+: 7/2^+[633]; 9/2^+: 9/2^+[624]$.}\\
\multicolumn{4}{l}{ $^{\rm b}$ An alternative configuration for
$K^\pi = 14^-$, which is denoted as Config1 in Fig.~\ref{fig:174Lumqp}(b).}\\
\multicolumn{4}{l}{ $^{\rm c}$ An alternative configuration for
$K^\pi = 15^+$, which is denoted as Config2 in Fig.~\ref{fig:174Lumqp}(c).}\\
\multicolumn{4}{l}{ $^{\rm d}$ An alternative configuration for
$K^\pi = 16^+$, which is denoted as Config3 in Fig.~\ref{fig:174Lumqp}(d).}\\
\multicolumn{4}{l}{ $^{\rm e}$ An alternative spin-parity assignment for the
state at 4710~keV.}
\end{tabular*}
\end{center}
\begin{multicols}{2}

The experimental kinematic MOI's for each band are extracted by
\begin{equation}
 \frac{J^{(1)}(I)}{\hbar^2} = \frac{2I+1}{E_{\gamma}(I+1\rightarrow I-1)} \ ,
\end{equation}
separately for each signature sequence within a rotational band.
The relation between the rotational frequency
$\omega$ and the angular momentum $I$ is
\begin{equation}
 \hbar\omega(I) = \frac{E_{\gamma}(I+1\rightarrow I-1)}{I_{x}(I+1)-I_{x}(I-1)} \ ,
\end{equation}
where $I_{x}(I)=\sqrt{(I+1/2)^{2}-K^{2}}$, $K$ is the projection of nuclear total angular
momentum along the symmetry $z$ axis of an axially symmetric nuclei.

Figure~\ref{fig:174Lu2qp} shows the experimental and calculated MOI's of
the 2-qp bands in ${}^{174}$Lu.
As we mentioned in Sec.~2, for a 2-qp band in a doubly-odd
nucleus there are four sequences of rotational levels.
The experimental MOI's of the high-$K$ ($K_> = \Omega_1+\Omega_2$) and low-$K$
($K_< = |\Omega_1-\Omega_2|$) bands are denoted by up and down triangles
(both full and empty), respectively.
The full and empty triangles denote the signature $\alpha=0$ and 1 bands, respectively.
The calculated MOI's by the PNC-CSM are denoted by
the dotted lines ($\alpha = 1$, $\alpha_\pi = +1/2, \alpha_\nu = +1/2$),
the dashed lines ($\alpha = 1$, $\alpha_\pi = -1/2, \alpha_\nu = -1/2$),
the short dashed lines($\alpha = 0$, $\alpha_\pi = -1/2, \alpha_\nu = +1/2$),
the solid lines ($\alpha = 0$, $\alpha_\pi = +1/2, \alpha_\nu = -1/2$).
It can be seen that nearly all the experimental MOI's are reproduced
by the PNC-CSM calculation quite well. An interesting thing is that
there is nearly no signature splitting in the $K_>$ or $K_<$ bands.
The splitting happens between the GM doublet.
The data of $K^\pi=7^+$ ($\pi9/2^-[514] \otimes \nu5/2^-[512]$) seems much
larger than it's GM partner band $K^\pi=2^+$.
One possible reason is that the spin-parity assignment and
the configuration assignment for this bands may be unreasonable because
there is nearly no signature splitting in both $\pi9/2^-[514]$ and $\nu5/2^-[512]$.

\end{multicols}
\begin{center}
\includegraphics[width=14cm]{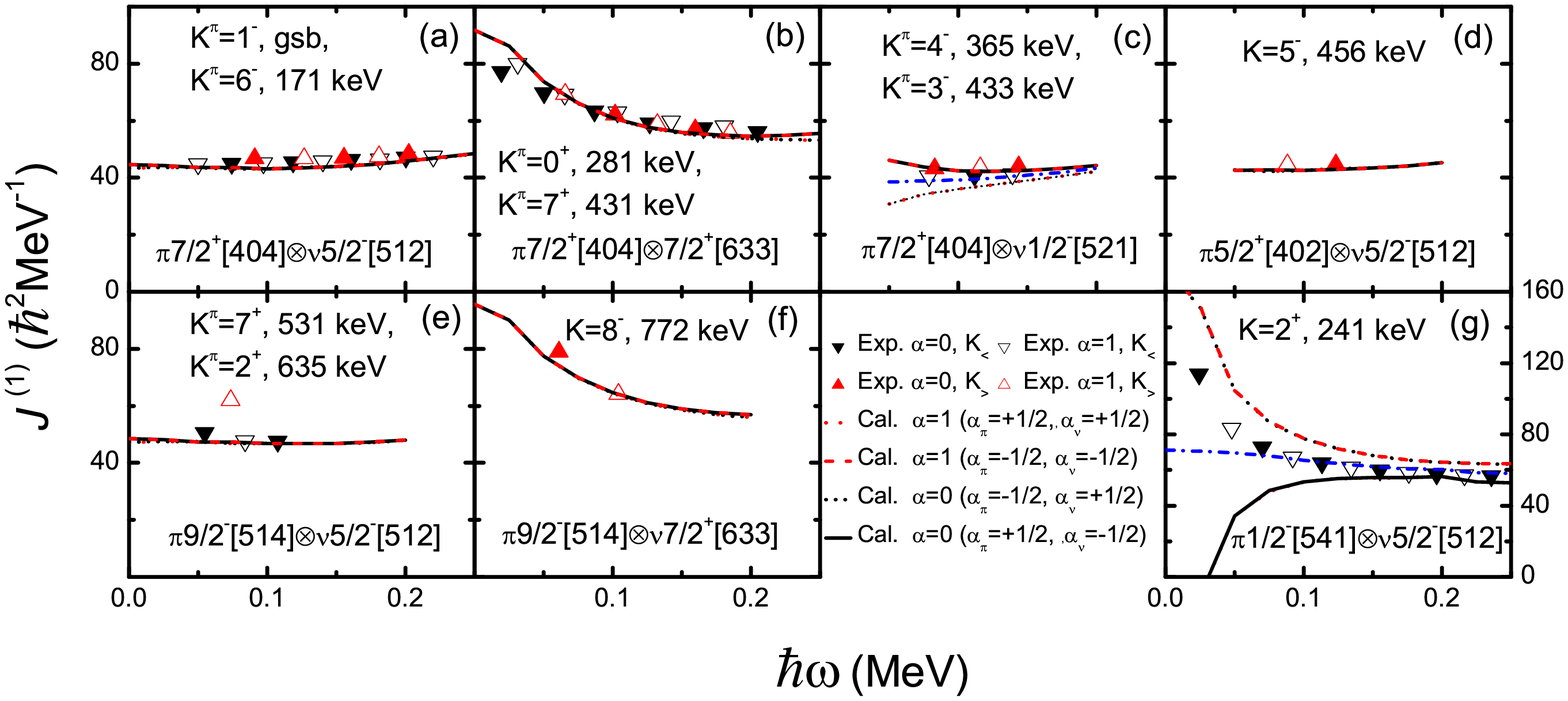}
\figcaption{  \label{fig:174Lu2qp}
(Color online)
The experimental and calculated MOI's $J^{(1)}$ of the 2-qp bands in ${}^{174}$Lu.
The experimental values are taken from Refs.~\citep{Bruder1987_NPA467-1,
Bruder1987_NPA474-518, Firestone1999, Dracoulis2006_PRL97-122501}.
The experimental MOI's of the high-$K$ ($K_> = \Omega_1+\Omega_2$)
and low-$K$ ($K_< = |\Omega_1-\Omega_2|$) bands are denoted by
up and down triangles (both full and empty), respectively.
The full and empty triangles denote the signature
$\alpha=0$ and 1 bands, respectively.
The calculated MOI's by the PNC-CSM are denoted by
the dotted lines       ($\alpha = 1$, $\alpha_\pi = +1/2, \alpha_\nu = +1/2$),
the dashed lines       ($\alpha = 1$, $\alpha_\pi = -1/2, \alpha_\nu = -1/2$),
the short dashed lines ($\alpha = 0$, $\alpha_\pi = -1/2, \alpha_\nu = +1/2$),
the solid lines        ($\alpha = 0$, $\alpha_\pi = +1/2, \alpha_\nu = -1/2$).
The blue dash dotted lines are the calculated results which are 50\% admixture
of the high-$K$ and low-$K$ states.
The pairing interaction strengths $G_p=0.33$~MeV and $G_n=0.30$~MeV
are determined by the experimental odd-even differences in nuclear binding energies.
}
\end{center}
\begin{multicols}{2}

It is well known that when one of the nucleons is in an $\Omega=1/2$ orbital, the GM doublet
has $\Delta K = 1$, accordingly the two bands are expected to be Coriolis admixed.
This effect can be very significant in the $K_<$ band, which has been identified in the
doubly-odd rare-earth nuclei~\cite{Oneil1972_NPA195-207, Jain1998_RMP70-843}.
In Fig.~\ref{fig:174Lu2qp} the blue dash dotted lines are the calculated results
which are 50\% admixing of the high-$K$ and low-$K$ states of a GM doublet.
As pointed out in Ref.~\cite{Oneil1972_NPA195-207}, the mixing in
$K^\pi = 2^+$ ($\pi1/2^-[541] \otimes \nu5/2^-[512]$) band is very strong,
which is almost 50\% mixing with it's high-$K$ GM partner band.
Our calculation shows that 50\% mixing can reproduce the data quite well in the higher
$\omega$ region, while in the very low $\omega$ region ($\hbar\omega < 0.05$~MeV),
the mixing is less prominent, which is about 30\%.
Another Coriolis mixing band is the $K^\pi=3^-$ ($7/2^+[404] \otimes \nu1/2^-[521]$) band.
This band also can be reproduced quite satisfactorily using 50\% mixing which
shows that in this GM doublet the mixing is as strong as that in
$K^\pi = 2^+$ ($\pi1/2^-[541] \otimes \nu5/2^-[512]$).

Figure~\ref{fig:174Lumqp} shows the experimental and calculated kinematic
MOI's of the multi-qp bands in ${}^{174}$Lu. All of these multi-qp isomers
are high-$K$ states. The experimental MOI's are denoted by the full up triangles
(signature $\alpha=0$) and the open up triangles (signature $\alpha=1$), respectively.
The calculated MOI's by the PNC-CSM are denoted by the solid lines
(signature $\alpha=0$) and the dotted lines (signature $\alpha=1$), respectively.
It can be seen that there is no signature splitting in these multi-qp isomers.

\begin{center}
\includegraphics[width=8cm]{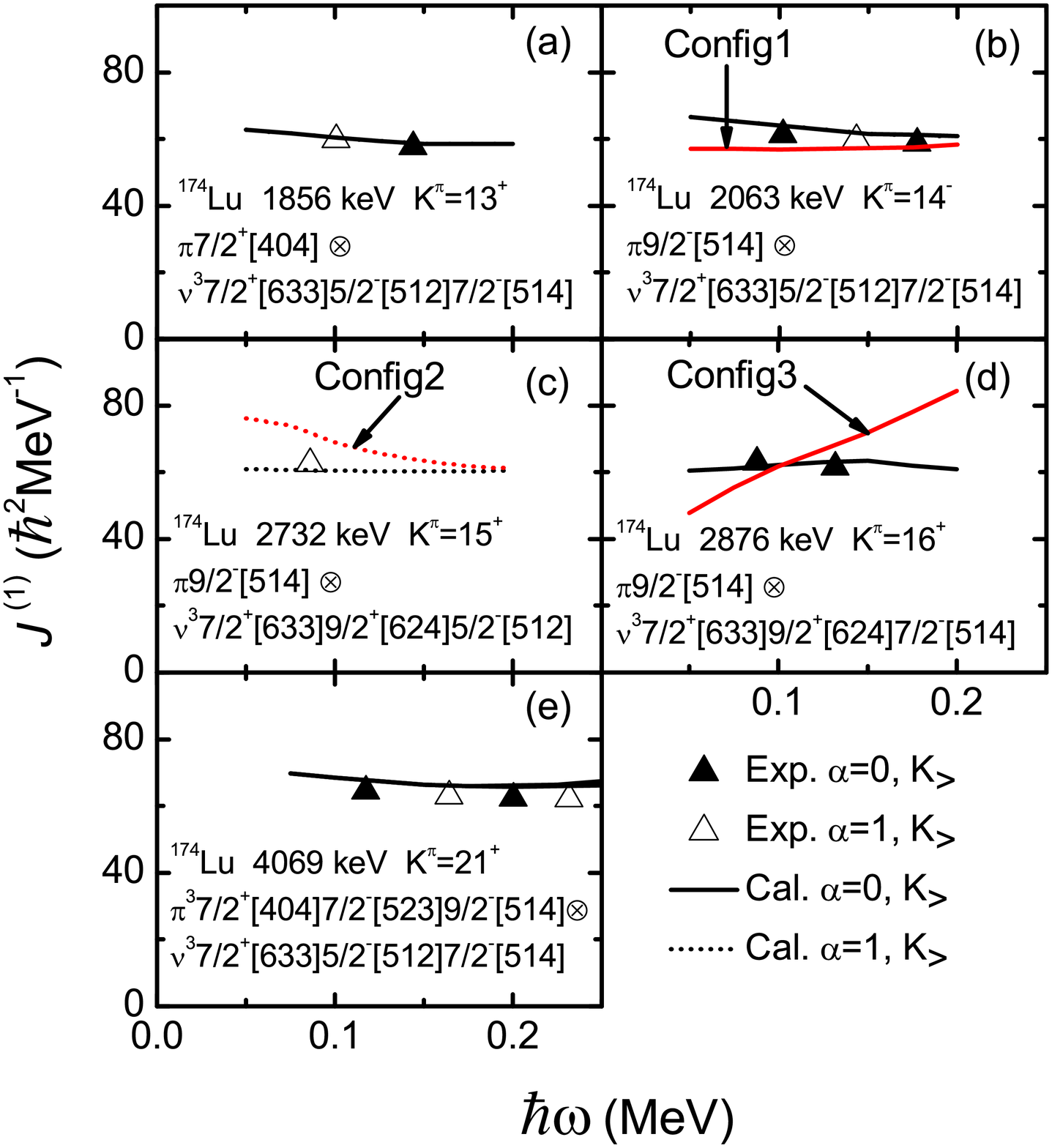}
\figcaption{  \label{fig:174Lumqp}
(Color online)
The experimental and calculated MOI's $J^{(1)}$ of the multi-qp bands in ${}^{174}$Lu.
The experimental values are taken from Ref.~\citep{Kondev2009_PRC80-014304}.
The experimental MOI's are denoted by full up triangles
(signature $\alpha=0$) and open up triangles (signature $\alpha=1$), respectively.
The calculated MOI's by the PNC-CSM are denoted by the solid lines
(signature $\alpha=0$) and the dotted lines (signature $\alpha=1$), respectively.
Config1 denotes an alternative configuration $\pi 7/2^+[404] \otimes
\nu^3 7/2^+[633] 9/2^+[624] 5/2^-[512]$ for $K^\pi = 14^-$ band.
Config2 denotes an alternative configuration $\pi^3 7/2^+[404]
7/2^-[523] 9/2^-[514] \otimes \nu 7/2^+[633]$ for $K^\pi = 15^+$ band.
Config3 denotes an alternative configuration $\pi^3 7/2^+[404]
7/2^-[523] 9/2^-[514] \otimes \nu 9/2^+[624]$ for $K^\pi = 16^+$ band.
}
\end{center}

There are several possible configuration assignments for these multi-qp states.
For $K^\pi=14^-$ band at 2063~keV, there are two possible configuration assignments.
In Ref.~\cite{Kondev2009_PRC80-014304}, the configuration is assigned as
$\pi 9/2^-[514] \otimes \nu^3 7/2^+[633] 5/2^-[512] 7/2^-[514]$ based on the $g_K$
value, the excitation energy and the alignment.
In Fig.~\ref{fig:174Lumqp}(b), MOI's of these two configurations are calculated
to compare with the data. An alternative configuration
$\pi 7/2^+[404] \otimes \nu^3 7/2^+[633] 9/2^+[624] 5/2^-[512]$
for $K^\pi = 14^-$ band is denoted as Config1 (red line).
Both the calculated MOI's of these two configurations are very close to the data.
The excitation energies of these two configurations from the PNC-CSM calculations
in Table~\ref{tab:174Lu} are also very close to each other.
So it is very hard to say which configuration is possible from our calculation.
For another two states $K^\pi=15^+$ at 2732~keV and $K^\pi=16^+$ at 2876~keV,
the configurations are chosen as
$\pi9/2^-[514] \otimes \nu^3 7/2^+[633]9/2^+[624]5/2^-[512]$ and
$\pi^3 7/2^+[404] 7/2^-[523] 9/2^-[514] \otimes \nu 9/2^+[624]$, respectively.
Figure~\ref{fig:174Lumqp}(c) and (d) shows that the experimental MOI's for
$K^\pi=15^+$ and $K^\pi=16^+$ bands can be reproduced quite well using these
two configuration assignments, respectively.
Other configuration assignments are also possible for these two states.
Config2 denotes an alternative configuration $\pi^3 7/2^+[404]
7/2^-[523] 9/2^-[514] \otimes \nu 7/2^+[633]$ for $K^\pi = 15^+$ band.
Config3 denotes an alternative configuration $\pi^3 7/2^+[404]
7/2^-[523] 9/2^-[514] \otimes \nu 9/2^+[624]$ for $K^\pi = 16^+$ band.
Obviously, the calculated MOI's of these two possible configurations
(see Fig.~\ref{fig:174Lumqp}(c) and (d)) are not consistent with the data,
which is consistent with Ref.~\cite{Kondev2009_PRC80-014304}.
The configuration of the 6-qp bands $K^\pi=21^+$ at 4069~keV is assigned as
$\pi^3 7/2^+[404] 7/2^-[523] 9/2^-[514] \otimes 7/2^+[633] 5/2^-[512] 7/2^-[514]$.
In Fig.~\ref{fig:174Lumqp} the data can be well reproduced by this configuration,
which also confirms the configuration assignment for this state.

\subsection{${}^{176}$Lu}

\begin{center}
\includegraphics[width=8cm]{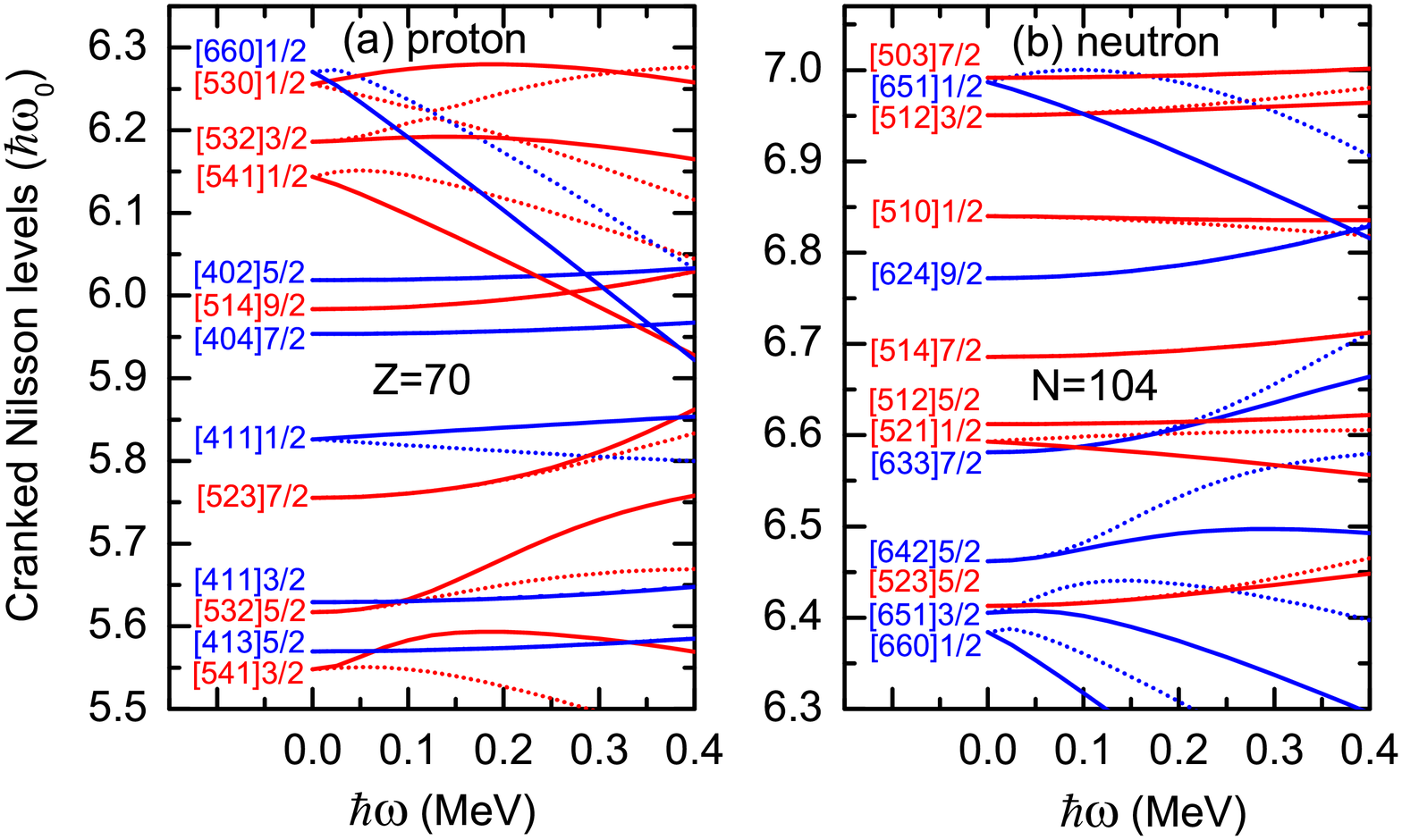}
\figcaption{  \label{fig:176LuNilsson}
(Color online) The cranked Nilsson levels near
the Fermi surface of ${}^{176}$Lu for (a) protons and (b) neutrons.
The deformation parameters ($\varepsilon_2 = 0.260, \varepsilon_4 = 0.0512$)
are taken from Lund systematics~\citep{Bengtsson1986_ADNDT35-15},
i.e., an average of the neighboring even-even Yb and Hf isotopes.
The proton (neutron) Nilsson level scheme of ${}^{176}$Lu is taken from that of
${}^{177}$Lu (${}^{177}$Hf) in Ref.~\citep{Zhang2009_PRC80-034313},
which is adopted to reproduced the bandhead energies of low-lying
1-quasiproton (1-quasineutron) bands of ${}^{177}$Lu (${}^{177}$Hf).
The positive (negative) parity levels are denoted by the blue (red) lines.
The signature $\alpha= +1/2$ ($\alpha = -1/2$) levels are denoted
by the solid (dotted) lines.
}
\end{center}

The cranked Nilsson levels near the Fermi surface of ${}^{176}$Lu are given
in Fig.~\ref{fig:176LuNilsson}.
The positive (negative) parity levels are denoted by the  blue (red) lines.
The signature $\alpha= +1/2$ ($\alpha = -1/2$) levels are denoted
by the solid (dotted) lines.
It should be noted that the proton level sequence
for ${}^{176}$Lu is slightly different from that of ${}^{174}$Lu.
The level sequence of $\pi9/2^-[514]$ and $\pi5/2^+[402]$ is conversed
in these two nuclei. The bandhead energies of the 2-qp states shown in
Table~\ref{tab:176Lu} confirm this. Most of the data are well reproduced by
the PNC-CSM calculations.

The experimental and calculated MOI's of the 2-qp bands in ${}^{176}$Lu
are shown in Fig.~\ref{fig:176Lu2qp}. Most of the data can be reproduced quite well
by the PNC-CSM calculations. Note that in ${}^{176}$Lu there are four GM doublets
which have $\Delta K = 1$. For $\pi1/2^-[541] \otimes \nu7/2^-[514]$ and
$\pi1/2^+[411] \otimes \nu7/2^-[514]$, the 50\% mixing of the low-$K$ states
with the high-$K$ states can reproduce the data very well.
The results show that in these two states the Coriolis mixing is also very strong.
But there are exceptions. Because the signature splitting of the neutron orbital
$\nu1/2^-[510]$ in the PNC-CSM calculation is very small, so the
Coriolis mixing in the $\pi7/2^+[404] \otimes \nu1/2^-[510]$
GM doublet can not be determined by the MOI's.
It is very interesting that the experimental MOI's of
$K^\pi = 3^-$ ($\pi7/2^+[404] \otimes \nu1/2^-[521]$) bands can be well reproduced
by the PNC-CSM calculations without Coriolis mixing.
We note the energy splitting of this GM doublet (50~keV) is much
smaller than others (usually about 100~keV). So the absence of the Coriolis mixing
of this state is due to the weaker proton-neutron residual interaction.

\begin{center}
\tabcaption{ \label{tab:176Lu}
The experimental and calculated 2-qp states in $^{176}$Lu.
The experimental values are taken from Ref.~\citep{Basunia2006_NDS107-791}.}
\footnotesize
\begin{tabular*}{80mm}{@{\extracolsep{\fill}}cccc}
\toprule
$K^\pi$   & Configuration                         &$E_{\rm Exp}$ & $E_{\rm Cal}$\\
~         & ~                                     &(keV)         & (keV)\\
\hline
$7^-, 0^-$& $\pi7/2^+[404] \otimes \nu7/2^-[514] $& 0, 123       & 0    \\
$1^+, 8^+$& $\pi9/2^-[514] \otimes \nu7/2^-[514] $& 194, 488     & 208  \\
$1^+, 8^+$& $\pi7/2^+[404] \otimes \nu9/2^+[624] $& 339, 425     & 409  \\
$1^-, 6^-$& $\pi5/2^+[402] \otimes \nu7/2^-[514] $& 386, 564     & 432  \\
$4^+, 3^+$& $\pi1/2^-[541] \otimes \nu7/2^-[514] $& 635, 734     & 1293 \\
$1^-, 6^-$& $\pi7/2^+[404] \otimes \nu5/2^-[512] $& 638, 766     & 518  \\
$3^-, 4^-$& $\pi7/2^+[404] \otimes \nu1/2^-[510] $& 658, 788     & 1014 \\
$4^-, 3^-$& $\pi1/2^+[411] \otimes \nu7/2^-[514] $& 723, 843     & 528  \\
$7^+, 2^+$& $\pi5/2^+[402] \otimes \nu9/2^+[624] $& 734, 866     & 841  \\
$0^-     $& $\pi9/2^-[514] \otimes \nu9/2^+[624] $& 780          & 617  \\
$4^-, 3^-$& $\pi7/2^+[404] \otimes \nu1/2^-[521] $& 908, 958     & 658  \\
$0^+, 7^+$& $\pi7/2^-[523] \otimes \nu7/2^-[514] $& 1057, 1274   & 1047 \\
\bottomrule
\end{tabular*}
\end{center}

\end{multicols}
\begin{center}
\includegraphics[width=14cm]{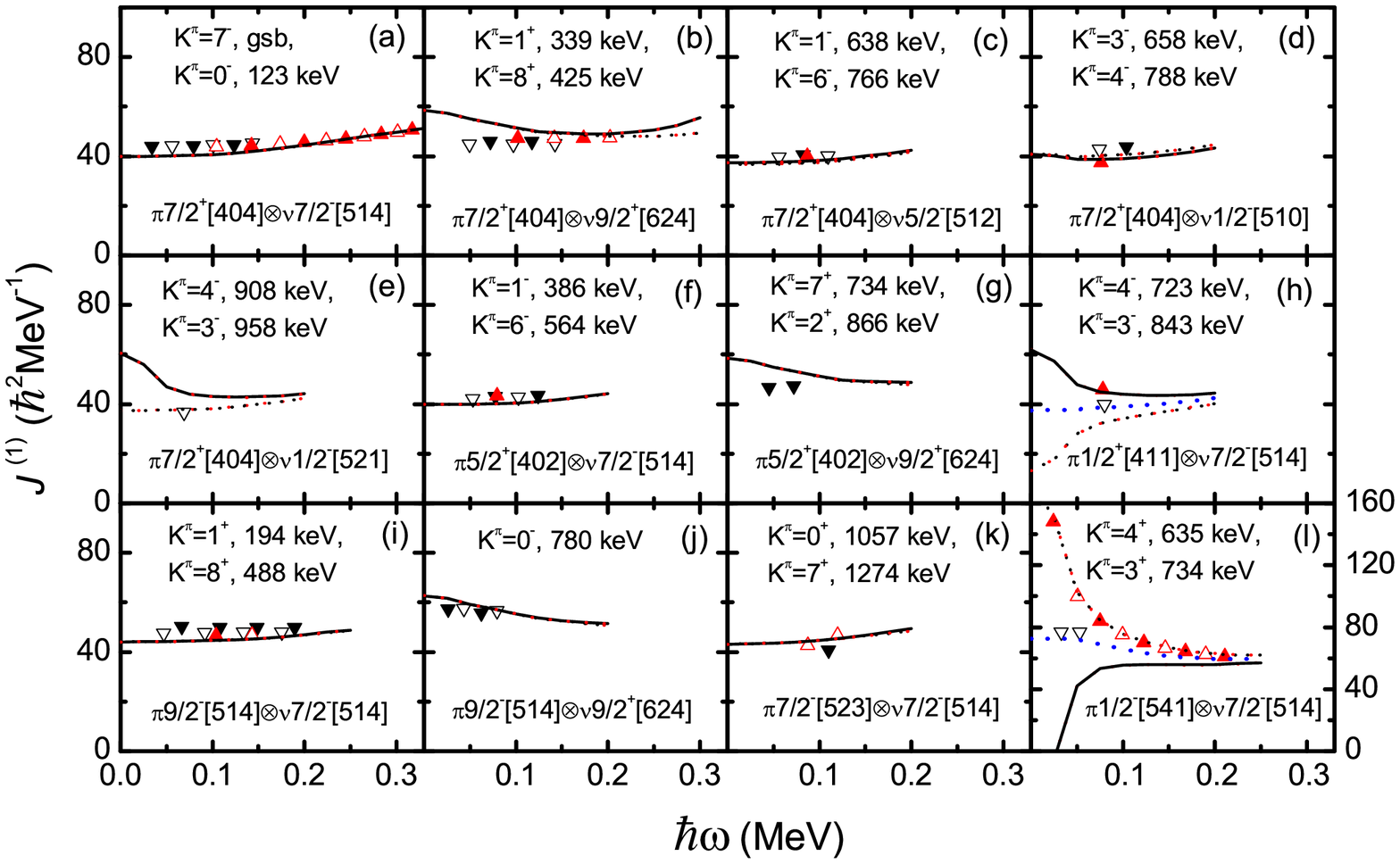}
\figcaption{ \label{fig:176Lu2qp}
(Color online)
The experimental and calculated MOI's $J^{(1)}$ of the 2-qp bands in ${}^{176}$Lu.
The experimental values are taken from Refs.~\citep{Basunia2006_NDS107-791,
Dracoulis2010_PRC81-011301R}.
The experimental MOI's of the high-$K$ ($K_> = \Omega_1 + \Omega_2$)
and low-$K$ ($K_< = |\Omega_1-\Omega_2|$) bands are denoted by
up and down triangles (both full and empty), respectively.
The full and empty triangles denote the signature
$\alpha=0$ and 1 bands, respectively.
The calculated MOI's by the PNC-CSM are denoted by
the dotted lines       ($\alpha = 1$, $\alpha_\pi = +1/2, \alpha_\nu = +1/2$),
the dashed lines       ($\alpha = 1$, $\alpha_\pi = -1/2, \alpha_\nu = -1/2$),
the short dashed lines ($\alpha = 0$, $\alpha_\pi = -1/2, \alpha_\nu = +1/2$),
the solid lines        ($\alpha = 0$, $\alpha_\pi = +1/2, \alpha_\nu = -1/2$).
The blue dash dotted lines are the calculated results which are 50\% admixture
of the high-$K$ and low-$K$ states.
The pairing interaction strengths $G_p=0.30$~MeV and $G_n=0.28$~MeV
are determined by the experimental odd-even differences in nuclear binding energies.
}
\end{center}
\begin{multicols}{2}

\section{Summary}{\label{Sec:summary}}
In summary, 2-qp bands and low-lying excited high-$K$ 4-, 6-, and 8-qp bands in
the doubly-odd ${}^{174, 176}$Lu are analyzed by using the cranked shell model with
the pairing correlations treated by a particle-number conserving method,
in which the blocking effects are taken into account exactly.
The proton and neutron Nilsson level schemes for ${}^{174, 176}$Lu are taken from the
adjacent odd-$A$ Lu and Hf isotopes, which are adopted to reproduce the experimental
bandhead energies of the one-quasiproton and one-quasineutron bands of these
odd-$A$ Lu and Hf nuclei, respectively.
The quasiparticle configurations are determined and the experimental bandhead energies
and the MOI's of these 2- and multi-qp bands are well reproduced by PNC-CSM calculations.
The Coriolis mixing of the low-$K$ ($K = |\Omega_1 - \Omega_2|$) 2-qp
band of the GM doublet involving $\Omega = 1/2$ is analyzed.
The possible configuration assignments of the multi-qp isomers in $^{174}$Lu are also
analyzed both from the bandhead energies and the MOI's.

\acknowledgments{The authors are grateful to Shan-Gui Zhou for fruitful discussions.}

\end{multicols}
\vspace{-3mm}
\centerline{\rule{80mm}{0.1pt}}
\vspace{2mm}
\begin{multicols}{2}


\end{multicols}
\clearpage
\end{document}